\documentclass[12pt]{iopart}

\usepackage{graphicx}
\usepackage{dcolumn}
\usepackage{bm}
\usepackage{hyperref}

\usepackage{multirow}%
\usepackage{amsmath,amssymb,amsfonts}%
\usepackage{tabularx,booktabs}
\usepackage{mathtools}
\usepackage{gensymb}

\newcommand{\KN}{\mathbin{\bigcirc\mspace{-15mu}\wedge\mspace{3mu}}}

\begin{document}

\title[On the Plebanski Formulation with Energy Momentum]{On the Plebanski Formulation with Energy Momentum}

\author{Jack C. M. Hughes$^{1,2}$, Joudy F. Jamal Beek$^{2}$ \& Fedor V. Kusmartsev$^{2}$}

\address{
  \small $^{1}$Department of Public Health and Epidemiology, College of Medicine and Health Sciences, Khalifa University, PO Box 127788, Abu Dhabi, United Arab Emirates\\
  \small $^{2}$Department of Physics, College of Engineering and Physical Sciences, Khalifa University, PO Box 127788, Abu Dhabi, United Arab Emirates}
\ead{jack.hughes@ku.ac.ae}
\vspace{10pt}
\begin{indented}
\item[]June 2026 
\end{indented}

\begin{abstract}
In Plebanski's formulation the coupling of matter is less direct than in the metric formulation since the energy-momentum tensor $T_{\mu\nu}$ is symmetric, while the Plebanski variables are naturally valued in the self-dual/anti-self-dual Hodge decomposition of 2-forms. An explicit construction of the Plebanski matter source $T^i$ is obtained by lifting the trace-free energy-momentum tensor $\hat T_{\mu\nu}$ into the $(1,1)$ component of the algebraic curvature space using the Kulkarni-Nomizu product, and then extracting its chiral components. This construction reproduces the definition for $T^i$ in terms of the self-dual basis $\Sigma^i$ and $\hat T_{\mu\nu}$ introduced by Krasnov. We also verify that the matter-coupled chiral field equations imply the usual conservation law $\nabla_{\mu}T^{\mu \nu}=0$ through the chiral Bianchi identity $d^A F^i=0$. As an application, the construction is applied to a spherically symmetric electromagnetic stress-energy tensor, where the anti-self-dual part of the matter-coupled Plebanski field equations yields the Reissner-Nordstr\"om-de Sitter solution. The result gives a systematic prescription for translating metric matter sources into the anti-self-dual source terms required by the Plebanski field equations.
\end{abstract}

\section{Introduction}
A great deal of technical progress in General Relativity (GR) has come from asking \textit{what kind of structure is needed to define the theory in four dimensions?} One way to approach this question is to shift emphasis on what the vacuum field equations are responsible for imposing. If 
\begin{equation}
    R_{\mu \nu} - \frac{1}{2} R g_{\mu \nu} + \Lambda g_{\mu \nu} = 0,
\end{equation}
then the manifold must be \textit{Einstein} \cite{besse_einstein_1987},
\begin{equation}\label{Einstein}
    R_{\mu \nu} = \Lambda g_{\mu \nu}.
\end{equation}
Consequently, one can develop mechanisms for generating Einstein metrics to `define' gravity \cite{fine_gauge_2016}.  This includes, for instance, holonomy reductions \cite{hall_symmetries_2004} (at least in exotic sectors) and the twistor correspondence \cite{penrose_spinor_1960, mason_integrability_1996}, which identifies \textit{self-dual} Einstein metrics with certain families of holomorphic curves via algebraic geometry \cite{besse_einstein_1987, krasnov_formulations_2020}. $G_2$ structures in seven dimensions are also known to generate Einstein metrics \cite{hitchin_stable_2001}, and it has been shown that the dimensional reduction of certain topological theories over SU(2) result in four dimensional Einstein metrics, with the cosmological constant emerging as the diameter of $S^3$ \cite{krasnov_general_2017, herfray_4d_2018} (similar results hold for three dimensions \cite{herfray_topological_2017}). In each signature, this provides the origin of the Urbantke metric \cite{urbantke_integrability_1984, krasnov_lorentzian_2023}. In addition, there exist classification schemes for (Euclidean) Einstein manifolds in terms of Yang-Mills instantons, drawing interesting connections to gauge theory \cite{park_anatomy_2022, chung_algebraic_2023, ho_einstein_2025, park_explicit_2024, lee_efficient_2011}. Einstein spaces are also known to act as attractors for the 'Einstein flow', or late time cosmological manifolds compatible with the observed homogeneity and isotropy of the Universe \cite{moncrief_could_2019, moncrief_einstein_2022, kouneiher_einstein_2015}. 

Parallel to this constructivism approach is the actual modification of the variables defining the gravitational action \cite{krasnov_formulations_2020, mielke_geometrodynamics_2017, gockeler_differential_1987, blagojevic_gauge_2012}. In Einstein's formulation we have a manifold equipped with a metric and metric-compatible connection. Of primary interest then is the tangent bundle $T\mathcal{M}$, with the symmetric 2-tensors (despite being reducible under the action of the orthogonal group \cite{besse_einstein_1987}) holding a privileged position as a consequence of the field equations being defined over this space. In this language, GR feels far removed from the foundations of the gauge theories as there is no sense in which the connection coefficients $\Gamma^{\mu}_{\;\rho \sigma}$ can be regarded as elements of a principal connection. Even if one elects to work with ``frame fields" $V^a_{\mu}(x)$ that define locally orthogonal inertial reference frames , the tangent bundle still takes precedent as the orthogonal frame bundle is derived from it \cite{gockeler_differential_1987, aldrovandi_teleparallel_2013}.

In contrast, Cartan's description of differential geometry encodes the structure of $\mathcal{M}$ into a set of differential forms assuming values in an abstract vector space $V$ \cite{krasnov_formulations_2020, shaw_general_2026}. This is made possible by the introduction of a new distinct vector bundle $E$ and the \textit{solder form} $e$ which generates the isomorphism 
\begin{equation}
    e \colon \; TM \to E.
\end{equation}
Any canonical structure defined on the fibers $V$ (via a reduction of the structure group \cite{harvey_spinors_1990}) may be pulled-back through this soldering map to an equivalent structure on $T\mathcal{M}$. The presence of the solder form is the center about which gauge theory and gravitation diverge; although the gauge connection (as a Lie algebra valued 1-form) establishes a similar mapping, objects which are pulled-back (e.g. a Killing form) will not remain gauge invariant since the connection transforms inhomogeneously. The standard Einstein-Cartan theory lets $V$ take the form of a Minkowski vector space, with the Christoffel connection being the pull-back of the principal-$\text{SO}(1,3)$ connection $\mathcal{A}$ on $E$, such that the gravitational action assumes the form \cite{krasnov_formulations_2020, gasperini_theory_2017}
\begin{equation}\label{EC_Action}
    S_{\text{EC}}[e, \mathcal{A}] = \frac{1}{32\pi G} \int \epsilon_{abcd} e^a \wedge e^b \wedge \bigg( R^{cd}(\mathcal{A}) - \frac{\Lambda}{6} e^c \wedge e^d\bigg). 
\end{equation}

The introduction of the abstract vector space $V$ opens the door for myriad GR formulations that each specify a different space $V$ of interest. For every choice of $V$, the nature of the respective soldering form changes and with it the algebraic structure of gravitation. This is exactly what occurs in the modified BF-theories \cite{celada_bf_2016, baez_introduction_1999, montesinos_diffeomorphism-invariant_2023, cattaneo_gravity_2024, freidel_relations_2012}. Since the Einstein-Cartan action (\ref{EC_Action}) involves the solder-form only in the 2-form combination $B^{ab} = (e\wedge e)^{ab}$, one can instead solder fiber-wise the bi-vector algebra
\begin{equation}
    B\colon T\mathcal{M}\otimes T\mathcal{M} \to \Lambda^2(\mathbb{R}^{1,3}). 
\end{equation}
However, as a vector space this is isomorphic to the Lie algebra of the Lorentz group:
\begin{equation}\label{B_Map}
    B\colon T\mathcal{M}\otimes T\mathcal{M} \to \mathfrak{so}(1,3). 
\end{equation}
The $\mathfrak{so}(1,3)$ canonical Killing form can be pulled-back through (\ref{B_Map}) to define an inner-product on the 2-forms. This is the formal foundation of the BF-Lagrangian
\begin{equation}\label{BF}
    S_{\text{BF}}[B, \mathcal{A}, \Psi] = \frac{1}{16\pi G} \int \text{Tr} \bigg(B \wedge R(\mathcal{A}) - \frac{1}{2}\bigg[\Psi + \frac{\Lambda}{6} \star\bigg] B \wedge B\bigg).
\end{equation}
Here $\Psi$ is a (traceless) Lagrange multiplier that ensures the resulting $B$ field is simple as a 2-form \cite{freidel_bf_1999}. The necessary existence of such constraints for the reproduction of GR is precisely the origin of non-trivial gravitational degrees of freedom.

There are other possibilities for the structure of $V$ beyond those that motivate a BF theory. $V$ can for instance be constructed as a coset of Lie algebras, $V \cong \mathfrak{g/h}$ where $\mathfrak{h}$ is a Lie sub-algebra of $\mathfrak{g}$ \cite{sharpe_differential_2000}. This is interesting, since at the bundle level (under the appropriate conditions) this can be interpreted as modeling $\mathcal{M}$ locally on a symmetric space $G/H$ \cite{wise_macdowell-mansouri_2010}. Such \textit{Cartan geometries} are the natural generalization of soldering to situations in which both the solder-form and Lorentz connection can be realized as two elements of a single \textit{Cartan connection}. Both Poincar\'e gauge theory and MacDowell-Mansouri fall into this class \cite{blagojevic_gauge_2012, wise_macdowell-mansouri_2010, randono_gauge_2010, thibaut_gravity_2026, krasnov_gravity_2018}, while in three dimensions Cartan geometry can be considered as the background for the topological formulations of GR \cite{wise_symmetric_2009}. If one instead elects to work at the level of tractor bundles \cite{attard_tractors_2017, slovak_notes_2021, attard_conformal_2018}, then this idea can be generalized to have conformal spheres about points and we can encode the conformal equivalence class $[g]$ of $\mathcal{M}$ locally. One can then isolate Einstein metrics (and hence vacuum solutions of interest) within $[g]$ via the construction of parallel tractors, an idea that can be given a variational basis \cite{herfray_einstein_2022, bonezzi_gravity_2010}.

The difficulty with (\ref{BF}) is that in soldering $V \cong \mathfrak{so}(1,3)$ one has already assumed a large amount of internal structure. Not only does one have a Killing form (constructed from the Minkowski metric $\eta_{ab}$), but one also has the invariant bi-linear $\epsilon_{abcd}$ that defines the Hodge dual on fibers (and is the origin of the topological Holst term \cite{holst_barberos_1996, rovelli_covariant_2014}). Once simplicity is enforced (through $\Psi$), these pull-back to their respective structures on $\mathcal{M}$. Again, these features can be traced back to the fact that (complexified) $\mathfrak{so}(1,3)$ is \textit{semi}-simple \cite{krasnov_formulations_2020, woit_euclidean_2021},
\begin{equation}\label{Lie_Alg_Decomp}
    \mathfrak{so}_{\mathbb{C}}(1,3) \cong \mathfrak{sl}(2,\mathbb{C}) \oplus \overline{\mathfrak{sl}(2,\mathbb{C})}. 
\end{equation}
A different approach then, unique to four dimensions, is to leverage the chiral structure of the Lorentz algebra (\ref{Lie_Alg_Decomp}) to our advantage. A truly remarkable result is that if we let all 2-forms in the action (\ref{EC_Action}) take values in only \textit{one half} of the semi-simple decomposition (\ref{Lie_Alg_Decomp}) - denoted here by subscript $+$ for self-dual - then  
\begin{equation}\label{Chiral_Action}
    S_{\text{Chiral}} [e,\mathcal{A}] = \frac{i}{8\pi G}\int (e^I \wedge e^J)_+ \wedge \bigg( R_{IJ}(\mathcal{A}_+) - \frac{\Lambda}{6} (e_I \wedge e_J)_+ \bigg)
\end{equation}
is equivalent to (\ref{EC_Action}) up to a topological (namely Holst) term \cite{krasnov_formulations_2020, atiyah_self-duality_1978, capovilla_self-dual_1991}. Rewritten in terms of spin variables, this is the well-known form \cite{carmeli_classical_2001, carmeli_reformulation_1977, samuel_lagrangian_1987}
\begin{equation}\label{Spin_Theory}    S_{\text{Spin}}[\sigma, A] = \frac{i}{32\pi G} \int \sigma^{A}_{\; \dot{A}} \wedge \sigma^{B \dot{A}} \wedge F_{AB}(A) - \frac{\Lambda}{12} \sigma^A_{\;\dot{A}} \wedge \sigma^{B\dot{A}} \wedge \sigma_{A}^{\;\dot{B}} \wedge \sigma_{B \dot{B}},
\end{equation}
where $\sigma_{A\dot{A}}$ is the spin solder form and $A$ is the spin connection. The Plebanski formalism is the construction of the (modified) chiral BF theory for (\ref{Spin_Theory}), in complete analogy with the BF transition from the Einstein-Cartan action (\ref{EC_Action}) to (\ref{BF}) \cite{capovilla_self-dual_1991, krasnov_plebanski_2011, capovilla_general_1989, pietri_so4_1999}:
\begin{equation}\label{Pleb_Action}
    S_{\text{Pleb}}[\Sigma, A, \Psi] = -\frac{i}{8\pi G} \int \Sigma^i \wedge F_i(A) - \frac{1}{2}\bigg( \Psi_{ij} + \frac{\Lambda}{3} \delta_{ij}\bigg) \Sigma^i \wedge \Sigma^j.
\end{equation}
The strength of Plebanski lies in its `minimalist' input: via the Urbanteke construction \cite{urbantke_integrability_1984} one can induce a spacetime metric from the constrained triple of Lie-algebra-valued 2-forms $\Sigma^i$, together with a unique torsion-free chiral connection $A^i$ relative to this basis. These objects encode GR through the field equation in $\Sigma^i$ that demands the curvature of $A^i$ be purely self-dual as a 2-form \cite{krasnov_formulations_2020}:
\begin{equation}\label{Pleb_Field_Eqn}
    F_i = \bigg(\Psi_{ij} + \frac{\Lambda}{3}\delta_{ij}\bigg) \Sigma^j.
\end{equation}
Unlike in (\ref{BF}), for Plebanski the (traceless) Lagrange multiplier field $\Psi_{ij}$ serves an active role on-shell as the components of the self-dual Weyl tensor. 

However despite all of its attractive features, in Plebanski the introduction of energy-momentum is not particularly transparent. The issue stems from the fact that in the metric formalism $T_{\mu \nu}$ is a symmetric 2-tensor, with the important property that the field equations enforce the conservation of $T_{\mu \nu}$ via the contracted Bianchi identity. In Einstein-Cartan theory (\ref{EC_Action}) one instead introduces a vector-valued 3-form to play this role, with the invariance of the action under diffeomorphisms giving now the associated conservation \cite{mielke_geometrodynamics_2017, gockeler_differential_1987}. Neither of these approaches lends themselves naturally to a 2-form representation, especially when chirality is involved. To couple matter to the Plebanski action additional Lagrange multipliers are typically required which subsequently have to be eliminated through the field equations \cite{capovilla_self-dual_1991}. For an arbitrary energy-momentum tensor of interest in (say) astrophysics, this is neither economical nor digestible relative to the metric formalism \cite{krasnov_plebanski_2011, tennie_consistent_2010}. 

On the other hand the influence of non-vanishing energy-momentum on the Plebanski field equation (\ref{Pleb_Field_Eqn}) is immediate from representation theory. The pure-trace element of $T_{\mu\nu}$ simply shifts the cosmological constant and hence must be a self-dual contribution, while the trace-free component acts as a source for the trace-free Ricci tensor which prevents the Einstein condition (\ref{Einstein}) and thus introduces a \textit{anti}-self-dual component to the curvature $F_i$, i.e. \cite{krasnov_formulations_2020, krasnov_plebanski_2011}
\begin{equation} \label{PFET}
    F_i = \left(\Psi_{ij} + \frac{1}{3}(\Lambda - 2 \pi G T)\delta_{ij} \right) \Sigma^j + 8 \pi G T_i,
\end{equation}
where $T_i$ is a anti-self-dual 2-form representing the trace-free energy-momentum content of matter.

Two questions of importance immediately follow the introduction of (\ref{PFET}):
\begin{enumerate}
    \item Given some $T_{\mu \nu}$, how does one construct the anti-self-dual 2-form $T_i$?
    \item Given such a $T_i$, does energy-momentum conservation follow from (\ref{PFET}) via the Bianchi identity in $F_i$?
\end{enumerate}
In this work, we construct $T_i$ via the self-dual projection of the Kulkarni-Nomizu product \cite{besse_einstein_1987, hughes_encoding_2026} between the trace-free energy-momentum tensor $\hat{T}$ and the metric, namely
\begin{equation}\label{Main_Form}
    T^i = 4i\big[P^+ (\hat{T} \KN g)\big]^{0i}_{\quad \mu \nu} dx^{\mu} \wedge dx^{\nu}.
\end{equation}
This again follows from simple representation theory of the special orthogonal group in four dimensions. (\ref{Main_Form}) is an economical formula that proves simple to implement on standard tensor algebra software (e.g. SageManifolds \cite{gourgoulhon_tensor_2015, joudyfjb_joudyfjbplebanski-formulation--energy-momentum_2026}). We also show that component-wise, this formula reproduces that known from Krasnov \cite{krasnov_formulations_2020}, 
\begin{equation}\label{EM_Pleb}
    \big(4i\big[P^+ (\hat{T} \KN g)\big]^{0i}\big)_{\mu \nu} = \Sigma^{i \; \; \rho}_{[\mu} \hat{T}_{\nu]\rho}.
\end{equation}

Using this, we can then demonstrate that the Bianchi identity $d^A F_i = 0$ implies that the field equation (\ref{PFET}) reduces to the consevation of energy-momentum, $\nabla_{\mu} T^{\mu \nu} = 0$, which can be viewed as an internal consistency check on the matter formalism (\ref{Main_Form}) with (\ref{EM_Pleb}). From this perspective, one also sees where GR and unimodular gravity diverge in the content of their theories \cite{montesinos_diffeomorphism-invariant_2023, hughes_encoding_2026, gielen_unimodular_2024, montesinos_trace-free_2025}: in  GR, the field equations say that curvature contains both pure self-dual and pure anti-self-dual 2-form data, while in the unimodular theory only the mixed sectors are accessible. This is why intuitively conservation must be \textit{assumed} to recover the missing sector, in contrast to following from the Bianchi identity.

Finally, we consider a non-trivial example of static spherically symmetric electro-vacuum, in which the formula (\ref{Main_Form}) takes on a special structure which allows the Reissner-Nordstrom-de Sitter solution to be derived.

\section{A Brief Review of the Plebanski Formalism}
The Plebanski formalism treats GR as a modified BF theory \cite{celada_bf_2016}, where the equations of motion in the Lagrange multiplier and connection enforce certain constraints that ensure we recover the appropriate structures for the chiral formalism (\ref{Chiral_Action})-(\ref{Spin_Theory}). For a complete review, see \cite{krasnov_formulations_2020, capovilla_self-dual_1991, krasnov_plebanski_2011}. In the following, we fix spacetime indices $\mu, \nu, \cdots$ via $\epsilon^{0123} = -1$ and reserve Latin indices $i,j,k$ for spatial/$\mathbb{C}^3$ directions.

As a consequence of the Lie algebra splitting (\ref{Lie_Alg_Decomp}), the space of 2-forms $\Lambda^2$ is reducible and (in spacetime signature) decomposes into two orthogonal three-dimensional complex vector spaces that are conjugates of eachother \cite{besse_einstein_1987, krasnov_formulations_2020, hughes_encoding_2026, hughes_warped_2026},
\begin{equation}\label{Lambda_Split}
    \Lambda^2 = \Lambda^+ \oplus \Lambda^-.
\end{equation}
These subspaces correspond to the eigenspaces of the Hodge dual, which squares to minus the identity on 2-forms. Crucially, the Hodge dual requires a metric for its introduction and thus this splitting is metric dependent (while in contrast the volume form requires only an orientation). 

The idea of Plebanski is to do-away with this metric dependence and instead induce it from a variational soldering procedure, similar to how in Einstein-Cartan theory the spacetime metric is the pull-back (through the solder-form) of a Lorentzian inner-product \cite{krasnov_formulations_2020, gockeler_differential_1987}. One begins from an oriented smooth 4-manifold and constructs the chiral solder form $\Sigma^i$ defining an \textit{arbitrary} three-dimensional complex subspace $\Tilde{\Lambda} \subset \Lambda^2$ via the fiber-wise isomorphism
\begin{equation}
    \Sigma^i \colon \mathbb{C}^3 \to  \Tilde{\Lambda} \subset \Lambda^2.
\end{equation}
The inner-product $\delta^{ij}$ on $\mathbb{C}^3$ will be puled-back through $\Sigma$ to provide a metric on $\Tilde{\Lambda}$, which is identified with the restricted natural inner-product on $\Lambda^2$ to $\Tilde{\Lambda}$ provided by the orientation \cite{krasnov_formulations_2020}. On this bundle we introduce an $\text{SO}(3, \mathbb{C})$ connection $A^i$, and these structures together are the starting point for the formalism.

Since we are interested here in spacetime, it is important to note that one must implement reality constraints on $\Sigma$ by hand (although in principle this could also be implemented variationally at the cost of further Lagrange multipliers) \cite{krasnov_plebanski_2011}:
\begin{equation}
    \Sigma^i \wedge \overline{\Sigma^j} = 0, \quad \text{Re}\big(\Sigma^i \wedge \Sigma_i\big) = 0.
\end{equation}

Beginning from (\ref{Pleb_Action}), variation in the Lagrange multiplier $\Psi$ (which is restricted to be traceless) results in the chiral soldering form $\Sigma^i \in \mathfrak{so}(3,\mathbb{C})$ being orthonormal modulo a top-form on $\mathcal{M}$,
\begin{equation}\label{Volume_Form}
    \frac{\delta S_p}{\delta \Psi^{ij}}\colon \quad \Sigma^i \wedge \Sigma^j \propto \delta^{ij} \epsilon_{\Sigma}.
\end{equation}
This has two important consequences \cite{krasnov_formulations_2020}:
\begin{enumerate}
    \item The first is that it ensures simplicity of the $\Sigma$'s, in the sense that they are guaranteed to assume the form
    \begin{equation} \label{Sig+}
        \Sigma^i_{\;\mu \nu}= i(e^0 \wedge e^i)_{\mu \nu}- \frac{1}{2}\epsilon^i_{\; jk}(e^j \wedge e^k)_{\mu \nu}.
    \end{equation}
    This is possible because the space of 2-forms $\Lambda^2$ admits a natural conformal metric structure via the wedge product. Restricting this structure to the subspace spanned by the $\Sigma$'s we can demand $\delta^{ij}$ pulls-back to this metric through the chiral solder form and with it enforce orthogonality in the basis.
    Note that this basis is, in particular, the self-dual projection of $e\wedge e$ appearing in the chiral action (\ref{Chiral_Action})
    \begin{equation}\label{defP+1}
        2i(P^+ e\wedge e)^{0i} = \Sigma^i, \quad P^{+ \; \alpha\beta}_{\mu \nu} = \frac{1}{2}\left( \delta^{\alpha}_{[\mu} \delta^{\beta}_{\nu]} - \frac{i}{2} \epsilon_{\mu \nu}^{\; \; \; \alpha \beta}\right)
    \end{equation}
    \item Via the Urbantke metric \cite{urbantke_integrability_1984}
    \begin{equation}
        g_{\Sigma}(u,v) \epsilon_{\Sigma} \sim \epsilon_{ijk} (i_u \Sigma^i) \wedge (i_v \Sigma^j) \wedge \Sigma^k
    \end{equation}
    one can select a representative from the conformal class $[g]$ relative to the volume form in (\ref{Volume_Form}) and with it build a pseudo-Riemannian metric for $\mathcal{M}$ whose signature is guaranteed via the simplicity property (\ref{Sig+}). The point is that under this identification the $\Sigma$'s span the subspace $\tilde{\Lambda} \cong\Lambda^+ \subset \Lambda^2$ that is induced thorough the metric $g_{\Sigma}$ (more generally its conformal class).
\end{enumerate}

Varying with respect to $A$ gives the condition that the $\text{SO}(3,\mathbb{C})$ connection be torsion free,
\begin{equation}\label{A_derivative}
    \frac{\delta S_P}{\delta A^i}\colon \quad d^A \Sigma^i = d\Sigma^i + \epsilon^i_{\; jk} A^j \wedge \Sigma^k =0.
\end{equation}
The goal of this constraint is to establish (provided orthogonality and simplicity are imposed) that $A$ is not a free variable but rather is uniquely determined via the triple $\Sigma^i$ \cite{krasnov_formulations_2020},
\begin{equation}
    A^i = A(\Sigma^i).
\end{equation}
This is entirely similar to how in Einstein-Cartan (\ref{EC_Action}), variation with respect to the Lorentz connection $\omega$ gives Cartan's first structure equation - $d^\omega e = 0$ - and results algebraically in a unique torsion free connection \cite{gockeler_differential_1987, hehl_gauge_2014, blagojevic_gauge_2012}. Again, the importance of this condition is clear when expressed in terms of the Lorentz connection (\ref{EC_Action}) 
\begin{equation}\label{A_i_P+}
    A^i = 2i(P_+ \mathcal{A})^{0i},
\end{equation}
or in words $A^i$ is the self-dual projection of the $\text{SO}(1,3)$ connection.

Finally, variation in the chiral solder form gives the Plebanski field equation\cite{plebanski_separation_1977}
\begin{equation}\label{Pleb_Field_Eqn}
    \frac{\delta S_{P}}{\delta \Sigma^i}\colon \quad F^i(A) = \bigg(\Psi_{ij} + \frac{\Lambda}{3} \delta_{ij} \bigg) \Sigma^i.
\end{equation}
Since $\Sigma^i$ spans the basis of $\Lambda^+$ (defined by the Urbantke metric to which it corresponds), this states that the curvature of $A^i$ is a self-dual 2-form. But again, since $A^i$ is the self-dual projection of the unique torsion-free Levi-Civita connection, this means that the manifold is Einstein \cite{atiyah_self-duality_1978} and hence gives vacuum GR.

Thus, (\ref{Pleb_Action}) provides a reformulation of vacuum GR in four dimensions. The initial data is an \textit{arbitrary} complex triple $\Sigma^i$ defining a 3-plane in $\Lambda^2$, together with an $\text{SO}(3,\mathbb{C})$ connection that allows us to differentiate sections of a $\mathbb{C}^3$ bundle. That this gives GR is a consequence primarily of the fact that this 3-plane identifies a conformal class $[g]$ with $\Sigma^i$ spanning $\Lambda^+_{[g]}$. The simplicity and reality conditions on $\Sigma^i$ identify a Lorentzian representative from this conformal class, and the torsion free condition on $A$ ultimately ties the differential topology of $\mathcal{M}$ to $\Lambda^+$.

\section{$T_i$ and the Kulkarni-Nomizu Product}
The introduction of energy momentum into the Plebanski formalism is a non-trivial problem \cite{capovilla_self-dual_1991, krasnov_plebanski_2011, tennie_consistent_2010, pillin_self-dual_1996}. As a symmetric 2-tensor, $T_{\mu \nu}$ is reducible under the action of the orthogonal group into its trace $T$ and trace-free $\hat{T}_{\mu \nu}$ components \cite{besse_einstein_1987},
\begin{equation} \label{T}
    T_{\mu \nu} = \hat{T}_{\mu \nu} + \frac{T}{4} g_{\mu \nu}.
\end{equation}
On the one hand the trace of the energy-momentum tensor $T$ simply shifts the value of the cosmological constant in the Einstein field equations, and hence is a purely self-dual contribution to the Plebanski field equation (\ref{Pleb_Field_Eqn}),
\begin{equation}
    F^i(A) = \bigg(\Psi_{ij} + \frac{\Lambda - 2\pi G T}{3} \delta_{ij} \bigg) \Sigma^i.
\end{equation}
On the other hand, the trace-free component of the energy-momentum tensor acts as a source for the trace-free Ricci tensor, which is not a purely self-dual element of the Riemann tensors orthogonal decomposition in four dimensions \cite{hughes_encoding_2026}. Instead, it induces a purely anti-self-dual contribution to the curvature $F^i$, which spoils the Einstein property (\ref{Einstein}), in accordance with the Einstein equations. In particular, we require \cite{krasnov_formulations_2020, krasnov_plebanski_2011}
\begin{equation}
    F^i(A) = \bigg(\Psi_{ij} + \frac{\Lambda - 2\pi G T}{3} \delta_{ij} \bigg) \Sigma^i + 8\pi GT^i,
\end{equation}
where $T^i$ is anti-self-dual as a 2-form.

The question of interest is how to construct $T^i$. In principle this can be achieved from a variational procedure in which matter fields are appropriately coupled to the chiral solder form so as to respect the trace-free character of the self-dual Weyl tensor $\Psi^{ij}$. This is obviously favorable as far as the philosophy as Plebanski is concerned, and the couplings may be kept polynomial in $\Sigma$ in this way (even for spinors) \cite{capovilla_self-dual_1991}. It is not clear, however, how useful this observation is for energy-momentum tensors that are introduced phenomenologically within the literature in the metric formalism.

To this end, Krasnov \cite{krasnov_formulations_2020, krasnov_plebanski_2011} introduced the following definition for $T^i$:
\begin{equation} \label{PT}
    T^i \coloneqq  T^i_{\, \mu \nu} dx^{\mu} \wedge dx^{\nu} = \Sigma^{i \;\;\rho}_{\;[\mu} \hat{T}_{\nu] \rho} dx^{\mu} \wedge dx^{\nu},
\end{equation} 
One can verify that this is purely anti-self-dual by computing the projection onto $\Lambda^+$. To see this, note that from (\ref{defP+1}) we have the identity (see (\ref{prfP}))
\begin{equation} \label{P+}
    P^{+ \; \; \rho \sigma}_{\; \mu \nu} = \frac{1}{4}\Sigma^i_{\;\mu \nu} \Sigma_i^{\;\rho \sigma},
\end{equation}
from which it follows
\begin{equation}
    P^{+ \; \;  \rho \sigma}_{\; \mu \nu} T^i_{\; \rho \sigma} = 0
\end{equation}
via the algebra of the $\Sigma$'s \cite{krasnov_formulations_2020}. From the definition (\ref{PT}), it is not immediately obvious why this particular combination is the correct choice. Furthermore, it proves difficult to implement this formula using standard tensor algebra packages and to recover the known results for, say, the perfect fluid (\cite{krasnov_plebanski_2011}). This motivates the following approach based on a lifting of a given energy-momentum tensor to the algebraic space of curvature tensors \cite{hughes_encoding_2026}.  

\subsection{The Kulkarni-Nomizu Product}
Over the differential forms on $\mathcal{M}$, one can introduce a graded algebra $\Omega(\mathcal{M})$ by symmetrizing the $p$-forms \cite{besse_einstein_1987},
\begin{equation}
    \Omega(\mathcal{M}) = \sum_{p=0}^n S^2\big( \Lambda^p \;T^*\mathcal{M}\big).
\end{equation}
For example, $S^2(\Lambda^1 \;T^*\mathcal{M})$ is simply the symmetric 2-tensors (e.g. the Ricci tensor). Similarly, $S^2(\Lambda^2 \;T^*\mathcal{M})$ contains 4-tensors sharing the same pair symmetries of the Riemann tensor,
\begin{equation}\label{ReimSym}
    X_{\mu \nu \rho \sigma} = -X_{\nu \mu \rho \sigma} = -X_{\mu \nu \sigma \rho} = X_{\rho \sigma \mu \nu}, \quad X\in S^2(\Lambda^2 \;T^*\mathcal{M}). 
\end{equation}

There exists a natural bilinear product structure over $\Omega(\mathcal{M})$, defined by \cite{besse_einstein_1987}
\begin{gather}\label{KN_General}
    \KN\colon S^2(\Lambda^p \;T^*\mathcal{M}) \times S^2(\Lambda^q \;T^*\mathcal{M}) \to S^2(\Lambda^{p+q} \;T^*\mathcal{M}) \\
    (\alpha \circ \beta)  \KN  (\mu \circ \nu)  = (\alpha \wedge \mu) \circ (\beta \wedge \nu)
\end{gather}
where $\alpha, \beta \in \Lambda^p T^*\mathcal{M}$, $\mu, \nu \in \Lambda^q T^*\mathcal{M}$, and $\circ$ denotes the symmetric tensor product. When $p = q = 1$, this is the Kulkarni-Nomizu (KN) product \cite{kulkarni_bianchi_1972, nomizu_decomposition_1972} (which provides the basis for the Riemann tensors orthogonal decomposition \cite{hughes_encoding_2026})
\begin{equation}
    (A \KN B)_{\mu \nu \rho \sigma} = A_{\mu \rho}B_{\nu \sigma} - A_{\mu \sigma }B_{\nu \rho} - A_{\nu \rho}B_{\mu \sigma} + A_{\nu \sigma }B_{\mu \rho} \in S^2(\Lambda^2 \; T^*\mathcal{M}).
\end{equation}
This 4-tensor - in addition to the symmetries (\ref{ReimSym}) - satisfies the first Bianchi identity
\begin{equation}
    (A \KN B)_{\mu \nu \rho \sigma} + (A \KN B)_{\nu \rho \mu \sigma} + (A \KN B)_{\rho \mu \nu \sigma} = 0.
\end{equation}
The subspace of $S^2(\Lambda^2 \; T^*\mathcal{M})$ satisfying the first Bianchi identity is known as the \textit{Algebraic space of curvature tensors}, $\mathcal{K}$.

Although (\ref{KN_General}) is defined over the (symmetric products of) differential forms, under the identification
\begin{equation}
    \Lambda^1 T^* \mathcal{M} = T^* \mathcal{M},
\end{equation}
the KN-product may also be applied to symmetric 2-tensors in the same way 
\begin{equation}
    \KN\colon S^2 (T^*\mathcal{M}) \times S^2 (T^*\mathcal{M}) \to \mathcal{K}.
\end{equation}
The main object of interest in this paper is the KN-product of the metric $g_{\mu \nu}$ with the traceless energy-momentum tensor $\hat{T}_{\mu \nu}$, which is given by 
\begin{equation} \label{TKNg}
    (\hat{T} \KN g)_{\mu \nu \rho \sigma}= \frac{1}{2}(g_{\mu [\rho} \hat{T}_{\sigma]\nu} - g_{\nu [\rho} \hat{T}_{\sigma]\mu}).
\end{equation}
Elements of $\mathcal{K}$ constructed in this way from trace-free tensors are naturally identified with homomorphisms from $\Lambda^+$ to $\Lambda^{-}$ (and vice-versa, via transposing). This follows immediately from representation theory: labeling the irreducible representations of $\mathfrak{so}_{\mathbb{C}}(1,3)$ via spin indices $(j_1,j_2)$, we have \cite{hughes_encoding_2026}
\begin{equation} \label{SDtoASD}
    \text{Hom}(\Lambda^+, \Lambda^-) \cong (1,0) \otimes (0,1) \cong (1,1),
\end{equation}
with $(1,1)$ being symmetric trace-free 2-tensors, while the space $\mathcal{K}$ itself has the following irreducible decomposition \cite{besse_einstein_1987, ita_instanton_2015}
\begin{equation}
    \mathcal{K} \cong (2,0) \oplus (0,2) \oplus (1,1) \oplus (0,0).
\end{equation}

\subsection{Kulkarni–Nomizu Construction of the Plebanski Energy–Momentum $T_i$}
In Plebanski's formulation \cite{capovilla_self-dual_1991,  krasnov_plebanski_2011, plebanski_separation_1977}, one expresses the (vacuum) field equations in terms of the self-dual basis $\Sigma^i$ (\ref{Sig+}) and the curvature components adapted to that basis (\ref{Pleb_Field_Eqn}). Trace-free matter contributions induce homomorphisms from the self-dual basis into the anti-self-dual basis and (to recover GR) necessarily contribute to the Plebanski field equation via a purely anti-self-dual sector of the Lie algebra decomposition (\ref{Lie_Alg_Decomp}). In order to extract this element from $g \KN \hat{T}$ (which is formally a $6\times 6$ matrix in the self-dual ant-self-dual basis), one must project into the associated $\mathfrak{so}(3,\mathbb{C})$ indices following (\ref{defP+1}) and (\ref{A_i_P+}):
\begin{equation} \label{PTKN2}
    T^i= 4i[P^+(\hat{T} \KN g)]^{0i}_{\;\; \mu \nu} dx^{\mu} \wedge dx^{\nu}.
\end{equation}
This can easily be shown to be equivalent to Kasnov's definition in (\ref{PT}) by expanding the components of (\ref{PTKN2}) using (\ref{defP+1})
\begin{equation}
    4i[P^+(\hat{T} \KN g)]^{0i}_{\;\; \mu \nu}= 4i\big[\frac{1}{2}((e^0\wedge e^i)_{\rho\sigma}(\hat{T} \KN g)^{\rho \sigma}_{\; \; \; \mu \nu}-\frac{i}{2}\epsilon^{0i}_{\;\;jk}(e^j\wedge e^k)_{\rho\sigma}(\hat{T} \KN g)^{\rho \sigma}_{\; \; \; \mu \nu})\big],
\end{equation}
and using $\epsilon^{0ijk}=-\epsilon^{ijk}$ with the definition of the self dual basis (\ref{Sig+}), it gives
\begin{equation} \label{PTKN}
    T^i =  \Sigma^i_{\;\rho \sigma}(\hat{T} \KN g)^{\rho \sigma}_{\;\;\mu \nu} dx^{\mu} \wedge dx^{\nu}.
\end{equation} 
Then expanding $  \Sigma^i_{\;\rho \sigma}(\hat{T}\KN g )^{\rho \sigma}_{\;\;\mu \nu}$ using (\ref{TKNg}) into 
\begin{equation}
    \Sigma^i_{\;\rho \sigma}(\hat{T}\KN g )^{\rho \sigma}_{\;\;\mu \nu}=\frac{1}{4}(\Sigma^i_{\;\mu \sigma}\hat{T}_{\nu}^{\;\sigma}-\Sigma^i_{\;\nu \sigma}\hat{T}_{\mu}^{\;\sigma}-\Sigma^i_{\;\rho \mu}\hat{T}_{\nu}^{\;\rho}+\Sigma^i_{\; \rho \nu}\hat{T}_{\mu}^{\;\rho}),
\end{equation}
and using the antisymmetry of the spacetime indices of the self-dual basis simplifies it to $T^i_{\;\mu \nu}$. 
Using these alternative representations of $T^i$ allows us to understand some of its properties more intuitively. From (\ref{PTKN2}), $T^i$ essentially takes the self-dual projection $P^+$ on the first index pair of $\hat{T} \KN g$ which, in terms of the $6 \times 6$ matrix representation with respect to the splitting $\Lambda^2=\Lambda^+\oplus\Lambda^-$, restricts us to the first row block of the matrix decomposition \cite{krasnov_formulations_2020, hughes_encoding_2026, montesinos_trace-free_2025, hughes_warped_2026}. Since the only non-zero component in this row block lies in the anti-self-dual sector, it follows that $T^i$ is anti-self-dual. The same conclusion can be seen directly from (\ref{PTKN}). As shown in (\ref{SDtoASD}), the operator $\hat{T} \KN g$ maps self-dual objects into anti-self-dual objects and vice-versa. Thus, it clear that the result of the action of ($\hat{T} \KN g$) on the self-dual basis $\Sigma^i$ produces an anti-self dual 2-form, precisely $T^i$. Another advantage of using the KN-product to express $T^i$ is that the derivation of the conservation of energy-momentum is relatively straight-forward, as will be shown in the next section.

\section{Conservation of Energy-momentum in Plebanski}
One of the central motivations for the form of the Einstein field equations is that, via a contraction over the second Bianchi identity, the covariant derivative of the Einstein tensor vanishes which enforces the conservation of energy-momentum. As Lovelock has shown, the class of 2-tensors that are concomitant to the metric (together with its first and second derivatives) and whose covariant derivative vanishes are exhausted by linear combinations of the Einstein tensor and the metic in four dimensions \cite{lovelock_fourdimensionality_1972}. This in turn influences the structure of the gravitational action, with energy-momentum conservation equivalently following from diffeomorphism invariance \cite{gockeler_differential_1987, straumann_general_2013}. In contrast, the trace-free formulation of gravity \cite{ellis_trace-free_2014} is only equivalent to GR provided one \textit{assumes} energy-momentum conservation \cite{hughes_encoding_2026}. This is interesting because despite it being impossible to construct a diffeomorphism invariant action that leads to the unimodular theory in the metric formalism \cite{blanckenburg_trace-free_2026}, this is possible via modified BF theory in a manner analogous to the Plebanski formalism (\ref{Pleb_Action}) \cite{montesinos_diffeomorphism-invariant_2023, gielen_unimodular_2024, montesinos_trace-free_2025}. As shown in \cite{hughes_encoding_2026}, it is possible to recover the unimodular theory via equating both the Ricci and $\hat{T} \KN g$ contributions to $(1,1)$ irreducible representations in $\mathcal{K}$ (via forming commutators with the Hodge dual), but one is still required to fix an algebraic constraint that reproduces energy-momentum conservation on $\mathcal{K}$ by hand (which translates into the chiral Bianchi identity below). The Plebanski field equation with energy-momentum (\ref{Pleb_Field_Eqn}) however contains both trace and trace-free irreducible representations, so one expects that energy-momentum conservation should follow by direct application of the Bianchi identity on the $\text{SO}(3,\mathbb{C})$ curvature $F^i$. We verify this using (\ref{PTKN2}).

The second Bianchi identity states that the exterior covariant derivative (induced by a connection) of the curvature vanishes \cite{gockeler_differential_1987, gasperini_theory_2017, straumann_general_2013}
\begin{equation} \label{BI}
    d^AF_i =0. 
\end{equation}
One can verify this by noting the definition of this derivative (\ref{A_derivative}) and using the Jacobi identity in the totally antisymmetric symbols $\epsilon^{i}_{\; jk}$. Applying (\ref{BI}) to the field equation (\ref{Pleb_Field_Eqn}) and distributing the derivative results in
\begin{equation}
    (d^A \psi_{ij}) \Sigma^j + \Psi_{ij} d^A \Sigma^j + \frac{\Lambda}{3} d^A \Sigma_i - \frac{2\pi G}{3}(dT \wedge \Sigma_i + T d^A \Sigma_i) + 8\pi G d^A T_i = 0,
\end{equation}
where the second, third and fifth terms vanish due to that vanishing of torsion (\ref{A_derivative}), leaving us with 
\begin{equation} \label{F1}
    (d^A \Psi_{ij}) \Sigma^j - \frac{2\pi G}{3}dT \wedge \Sigma_i + 8\pi G d^A T_i = 0.
\end{equation}
Writing this equation in terms of the self-dual projection operator using (\ref{PTKN2}), 
\begin{equation}
   \Psi^{ij} \Sigma_j = 4i[P^+C]^{0i}_{\;\; \mu \nu} dx^{\mu} \wedge dx^{\nu}
\end{equation}
where $C_{\alpha \beta \gamma \delta}$ the Weyl tensor, and 
\begin{equation}
 \Sigma^i = 2i[P^+(e \wedge e)]^{0i}_{\;\; \mu \nu} dx^{\mu} \wedge dx^{\nu},
\end{equation}
results in 
\begin{equation} \label{F2}
    d^A[P^+C]_{0i} - \frac{\pi G}{3}dT \wedge [P^+(e \wedge e)]_{0i} + 8\pi G d^A[P^+(\hat{T} \KN g)]_{0i} = 0
\end{equation}
after factoring out $4i$.
We then extract the components of equation (\ref{F2}) by changing the covariant exterior derivative $d^A$ into the total covariant derivative $D_{\lambda}$ using 
\begin{equation}
    (d^A\omega^i)_{\lambda \mu_1 \dots \mu_p}=(p+1)\,D_{[\lambda}\,\omega^i_{\mu_1 \dots \mu_p]},
\end{equation}
where $\omega$ here is a Lie algebra valued $p$-form \cite{gockeler_differential_1987, gasperini_theory_2017}. This results in
\begin{equation} \label{F2.5}
    3(D_{[\lambda|}[P^+C]_{0i|\mu \nu]}- \pi G \nabla_{[\lambda|}T[P^+(e \wedge e)]_{0i|\mu \nu]}+24\pi G D_{[\lambda|}[P^+(\hat{T} \KN g)]_{0i|\mu \nu]} =0.
\end{equation}

Now, we take the hodge dual of equation (\ref{F2.5}) which is defined on 3-forms as 
\begin{equation}
   (\star X^i)_\gamma = \frac{1}{3!} \epsilon_{\gamma}^{\;\lambda \mu \nu}X^i_{\;\lambda \mu \nu},
\end{equation}
where $X^i$ is a Lie algebra valued 3-form, and $\epsilon_{\gamma}^{\;\lambda \mu \nu}$ is the Levi-Civita tensor. Then, using the identity (proved in \ref{prfH}) 
\begin{equation} \label{HD}
    \frac{1}{3!} \epsilon_{\gamma}^{\;\lambda \mu \nu} (3\nabla_{[\lambda}Y_{\mu \nu]})= -\nabla^{\sigma}(\star Y)_{\sigma \gamma}
\end{equation}
with $Y$ as a 2-form, we reach the equation
\begin{equation} \label{F3}
 -\nabla^{\sigma}[P^+C]_{0i \sigma \gamma}+ \frac{\pi G}{3}\nabla^{\sigma}T[P^+(e \wedge e)]_{0i\sigma \gamma}+8\pi G \nabla^{\sigma}[P^+(\hat{T} \KN g)]_{0i \sigma \gamma} =0 
\end{equation}
after factoring out a common $i$ from the equation. Here we use the fact that the hodge dual of a 2 form $Y$ is $\star Y =\pm i Y $ where the positive result corresponds to self dual 2-forms and the negative one for anti-self dual 2-forms.
Noting that 
\begin{equation}
    (P^+Z)_{0i \rho \sigma} = e_0^{[\mu}e_i^{\nu]}P^{+ \; \alpha\beta}_{\mu \nu}Z_{\alpha \beta \rho \sigma},
\end{equation}
with $Z_{\alpha \beta \rho \sigma}$ as a Riemann tensor-like object (\ref{ReimSym}), equation (\ref{F3}) can be written as 
\begin{equation} \label{F5}
  e_0^{[\mu}e_i^{\nu]}P^{+ \; \alpha\beta}_{\mu \nu}\left[-\nabla^{\sigma}C_{\sigma \gamma \alpha \beta}+ \frac{2\pi G}{3}\nabla_{[\alpha}Tg_{\beta] \gamma}+8\pi G \nabla^{\sigma}(\hat{T} \KN g)_{\sigma \gamma \alpha \beta}\right] =0, 
\end{equation}
This equation says that only the $0i$-components of the LHS vanish. But since the self-dual projection of the bracketed expression in (\ref{F5}) is self-dual, it is completely determined by its $0i$-components (because generally, every $0i$-component is related to a $jk$-component via the hodge dual and if it is self-dual then the hodge dual gives itself up to a factor and so the $0i$-components $\propto$ $jk$-components, where $i \neq j \neq k$). Hence, the vanishing of its $0i$-components implies the vanishing of the full self-dual tensor
\begin{equation}
  P^{+ \; \alpha\beta}_{\mu \nu}\left[-\nabla^{\sigma}C_{\sigma \gamma \alpha \beta}+ \frac{2\pi G}{3}\nabla_{[\alpha}Tg_{\beta] \gamma}+8\pi G \nabla^{\sigma}(\hat{T} \KN g)_{\sigma \gamma \alpha \beta}\right] =0.
\end{equation}
We then use the definition of $(\hat{T} \KN g)_{\sigma \gamma \alpha \beta}$ in (\ref{TKNg}) and substitute for the trace-free energy-momentum tensor (\ref{T}):
\begin{equation} \label{F4}
    \begin{aligned}
    P^{+ \; \alpha\beta}_{\mu \nu}\Bigl(&-\nabla^{\sigma}C_{\sigma \gamma \alpha \beta}+ \frac{2\pi G}{3}\nabla_{[\alpha}Tg_{\beta] \gamma} \\
    &+ 4\pi G (\nabla_{[\alpha}T_{\beta] \gamma} - \frac{1}{4}\nabla_{[\alpha}Tg_{\beta] \gamma} - g_{\gamma [\alpha}\nabla^{\rho}T_{\beta] \rho} +\frac{1}{4}g_{\gamma [\alpha}\nabla_{\beta]}T) \Bigr)=0.   
    \end{aligned}
\end{equation}
Using 
\begin{equation}
    g_{\gamma [\alpha}\nabla_{\beta]}T = -\nabla_{[\alpha}Tg_{\beta] \gamma},
\end{equation}
if follows that equation (\ref{F4}) simplifies to 
\begin{equation}
    P^{+ \; \alpha\beta}_{\mu \nu}\left(-\nabla^{\sigma}C_{\sigma \gamma \alpha \beta}+ 4\pi G (\nabla_{[\alpha}T_{\beta] \gamma} - g_{\gamma [\alpha}\nabla^{\rho}T_{\beta] \rho}  - \frac{1}{3}\nabla_{[\alpha}Tg_{\beta] \gamma}) \right)=0. 
\end{equation}
Then, we expand $P^{+ \; \alpha\beta}_{\mu \nu}$ using the definition (\ref{P+}) which results in a real and an imaginary part individually equal to zero. Contracting with $g^{\gamma \nu}$ shows that the imaginary part vanishes identically while the real part gives
\begin{equation}
     \left(\delta^{\alpha}_{\mu}g^{\gamma \beta}- g^{\gamma \alpha}\delta^{\beta}_{\nu}\right) \left(-\nabla^{\sigma}C_{\sigma \gamma \alpha \beta}+ 4\pi G (\nabla_{[\alpha}T_{\beta] \gamma} - g_{\gamma [\alpha}\nabla^{\rho}T_{\beta] \rho}  - \frac{1}{3}\nabla_{[\alpha}Tg_{\beta] \gamma})\right)=0.  
\end{equation}
Expanding the brackets results in the Weyl related terms vanishing as the Weyl tensor is traceless, and simplifying the rest of the terms causes the $\nabla_{\mu}T$ terms to cancel, resulting in 
\begin{equation}
    8 \pi G \nabla^{\rho}T_{\rho \mu} =0.
\end{equation}

Thus, the traditional conservation of energy-momentum is enforced via the second Bianchi within the Plebanski formalism equipped with the appropriate anti-self-dual energy-momentum 2-form (\ref{PTKN2}).

\section{Electro-Vacuum}
As an example, we now use Plebanski's formulation as demonstrated by Krasnov in \cite{krasnov_plebanski_2011} with the definition of $T^i$ proposed in (\ref{PTKN}) to derive the Reissner-Nordstr\"om-de Sitter spacetime from the spherically symmetric ansatz \cite{straumann_general_2013}
\begin{equation}
    ds^2= -e^{2v(t,r)}dt^2+e^{2f(t,r)}dr^2+r^2 d\theta^2+r^2 \sin^2(\theta)d\phi^2,
\end{equation}
together with an electromagnetic 2-form $\tilde{F}$ generated by the connection $\tilde{A}$ for a spherical charge distribution of strength Q,
\begin{equation}
    \tilde{A}= - \frac{Q}{r}dt, \quad \quad \quad \tilde{F}= d\tilde{A} = -\frac{Q}{r^2}dt \wedge dr.
\end{equation}
We introduce the following orthonormal basis of 1-forms:
\begin{equation} \label{tetrad}
    e^t = e^{v(t,r)}dt, \quad \quad e^r = e^{f(t,r)}dr, \quad \quad e^{\theta}=rd\theta, \quad \quad e^{\phi}=r \sin(\theta) d\phi.
\end{equation}
Then use (\ref{tetrad}) to define our self-dual basis as in (\ref{Sig+})
\begin{equation}
    \begin{aligned}
    \Sigma^1 &= i\,e^{f(t,r)+v(t,r)}dt \wedge dr - r^2\sin(\theta)d\theta \wedge d\phi, \\ \Sigma^2 &= i\,re^{v(t,r)}dt \wedge d\theta           + r\,e^{f(t,r)}\sin(\theta)dr \wedge d\phi, \\ \Sigma^3 &=i\,re^{v(t,r)}\sin(\theta)dt \wedge d\phi- re^{f(t,r)}dr \wedge d\theta.
    \end{aligned}
\end{equation}
Notice that $\Sigma^2$ and $\Sigma^3$ are related via
\begin{equation}\label{Hodge_2d}
    \Sigma^3 = \star_{S^2} \Sigma^2,
\end{equation}
where $\star_{S^2}$ is the hodge dual defined on the 2-sphere ($d\theta, \sin (\theta) d\phi$). In other words, they are related by a $90 \degree$ rotation about the internal angular plane as the result of the spherical symmetry in the metric ansatz.
We now compute the associated connection $A^i$, which is defined by the equation of motion 
\begin{equation} \label{EoMinSig}
    d\Sigma^i + \epsilon^{ijk}A^j \wedge \Sigma^k=0,
\end{equation}
resulting from varying the Plebanski action with respect to the triple $\Sigma^i$. Equation (\ref{EoMinSig}) can be written as a system of 12 linear equations with 12 unknown $A^i$ components (each $A^i$ has 4 components and the index $i$ runs over $1,2,3$), and solving the system in this case results in 
\begin{equation}
    \begin{aligned}
A^1 &= i\,e^{-f(t,r)+v(t,r)}
      \frac{\partial v}{\partial r}\,dt
      + i\,e^{f(t,r)-v(t,r)}
      \frac{\partial f}{\partial t}\,dr
      + \cos(\theta)\,d\phi, \\
A^2 &= -e^{-f(t,r)}\sin(\theta)\,d\phi, \\
A^3 &= e^{-f(t,r)}\,d\theta.
\end{aligned}
\end{equation}
Next, the curvature $F^i$ is simply computed using 
\begin{equation}
    F^i = dA^i + \frac{1}{2} \epsilon^{ijk}A^j \wedge A^k,
\end{equation}
to give 
\begin{equation}
\begin{aligned}
F^1
&=
-e^{-(f+v)}\Bigg(
-i e^{2f}\left(\frac{\partial f}{\partial t}\right)^2
+i e^{2f}\frac{\partial f}{\partial t}\frac{\partial v}{\partial t}
-i e^{2f}\frac{\partial^2 f}{\partial t^2}
-i e^{2v}\frac{\partial f}{\partial r}\frac{\partial v}{\partial r}
+i e^{2v}\left(\frac{\partial v}{\partial r}\right)^2
\\
&+i e^{2v}\frac{\partial^2 v}{\partial r^2}
\Bigg)dt \wedge dr -\left(1-e^{-2f}\right)
\sin(\theta)\, d\theta \wedge d\phi ,
\\
F^2
&= -i e^{-2f+v} \frac{\partial v}{\partial r} dt \wedge d\theta +e^{-f}\sin(\theta) \frac{\partial f}{\partial t} dt \wedge d\phi -i e^{-v}
\frac{\partial f}{\partial t} dr \wedge d\theta
+e^{-f}\sin(\theta)
\frac{\partial f}{\partial r} dr \wedge d\phi ,
\\
F^3&=-e^{-f}
\frac{\partial f}{\partial t}dt \wedge d\theta
-i e^{-2f+v}\sin(\theta)
\frac{\partial v}{\partial r}dt \wedge d\phi
-e^{-f}
\frac{\partial f}{\partial r}dr \wedge d\theta
-i e^{-v}\sin(\theta)
\frac{\partial f}{\partial t}dr \wedge d\phi .
\end{aligned}
\end{equation}
Notice that the relationship between $\Sigma^2$ and $\Sigma^3$ (\ref{Hodge_2d}) manifests between $A^2$ and $A^3$, and concomitantly between $F^2$ and $F^3$. 
Now, we compute the Plebanski energy-momentum $T^i$ using (\ref{TKNg}) and (\ref{PTKN}) from the trace-free electromagnetic stress-energy tensor
\begin{equation}
    T_{\mu\nu}=\frac{1}{4\pi}\left(g^{\alpha\beta}\tilde{F}_{\alpha\mu}\tilde{F}_{\beta\nu} -\frac{1}{4}g_{\mu\nu}\tilde{F}_{\alpha\beta}\tilde{F}^{\alpha\beta}\right),
\end{equation}
giving
\begin{equation}
\begin{aligned}
T^1
&=
-\frac{iQ^{2}e^{-(f+v)}}{8\pi r^{4}}dt\wedge dr
-\frac{Q^{2}e^{-2(f+v)}\sin(\theta)}{8\pi r^{2}}d\theta\wedge d\phi,
\\
T^2&= 0,
\\
T^3&=0.
\end{aligned}
\end{equation}
It is convenient to extract the Einstein equations when $F^i$ and $T^i$ are written in terms of the self-dual $\Sigma^i$ and anti-self-dual 2-forms $\bar{\Sigma}^i=-(\Sigma^i)^*$ instead of the coordinate 2-forms:
\begin{equation}
\begin{alignedat}{2}
dt\wedge dr&=\frac{e^{-(f+v)}}{2i}(\bar{\Sigma}^1+\Sigma^1), \qquad &d\theta \wedge d\phi&=\frac{1}{2r^2 \sin(\theta)}(\bar{\Sigma}^1-\Sigma^1),
\\
dt\wedge d\theta&=\frac{e^{-v}}{2ir}(\bar{\Sigma}^2+\Sigma^2), \qquad &dr \wedge d\phi&= -\frac{e^{-f}}{2r \sin(\theta)}(\bar{\Sigma}^2-\Sigma^2),
\\
dt\wedge d\phi&=\frac{e^{-v}}{2ir\sin(\theta)}(\bar{\Sigma}^3+\Sigma^3), \qquad &dr \wedge d\theta&=\frac{e^{-f}}{2r}(\bar{\Sigma}^3+\Sigma^3).
\end{alignedat}
\end{equation}
Thus, $F^i$ takes the form 
\begin{equation}
\begin{aligned}
F^1
&=
-\frac{e^{-2(f+v)}}{2}\Bigg[\left(
-e^{2f}\left(\frac{\partial f}{\partial t}\right)^2
+e^{2f}\frac{\partial f}{\partial t}\frac{\partial v}{\partial t}
-e^{2f}\frac{\partial^2 f}{\partial t^2}
-e^{2v}\frac{\partial f}{\partial r}\frac{\partial v}{\partial r}
+e^{2v}\left(\frac{\partial v}{\partial r}\right)^2
+e^{2v}\frac{\partial^2 v}{\partial r^2}
\right)\\
&(\bar{\Sigma}^1+\Sigma^1) 
+\frac{e^{2v}}{r^2}\left(e^{2f}-1\right)(\bar{\Sigma}^1-\Sigma^1)\Bigg] ,
\\
F^2
&= -\frac{e^{-2f}}{2r} \Bigg(\frac{\partial v}{\partial r}(\bar{\Sigma}^2+\Sigma^2)
+\frac{\partial f}{\partial r} (\bar{\Sigma}^2-\Sigma^2) \Bigg)
-i\frac{e^{-(f+v)}}{r}\frac{\partial f}{\partial t}(\bar{\Sigma}^3),
\\
F^3
&= -\frac{e^{-2f}}{2r} \Bigg(\frac{\partial v}{\partial r}(\bar{\Sigma}^3+\Sigma^3)
+\frac{\partial f}{\partial r} (\bar{\Sigma}^3-\Sigma^3) \Bigg)
+i\frac{e^{-(f+v)}}{r}\frac{\partial f}{\partial t}(\bar{\Sigma}^2).
\end{aligned}
\end{equation}
While $T^i$ takes the following anti-self-dual form
\begin{equation}
\begin{aligned}
T^1
&=
-\frac{Q^{2}e^{-2(f+v)}}{16\pi r^{4}}(\bar{\Sigma}^1+\Sigma^1)
-\frac{Q^{2}e^{-2(f+v)}}{16\pi r^{4}}(\bar{\Sigma}^1-\Sigma^1) = -\frac{Q^{2}e^{-2(f+v)}}{8\pi r^{4}}\bar{\Sigma}^1,
\\
T^2&= 0,
\\
T^3&=0.
\end{aligned}
\end{equation}

The Plebanski equation (\ref{PFET}) can be understood as selecting the first row block of the Riemann tensor expressed in the self-dual/anti-self-dual block decomposition. In this form, the trace-free Ricci tensor is encoded in the off-diagonal self-dual/anti-self-dual blocks (\ref{SDtoASD}). These off-diagonal components contain the 9 trace-free equations of Einstein's field equations, while the remaining trace equation is treated separately. Since the curvature two-forms $F^i$ encode this first row block, their anti-self-dual components correspond precisely to the trace-free Ricci sector. Therefore, the trace-free Einstein equations are recovered by equating the anti-self-dual part of $F^i$ to the anti-self-dual Plebanski energy-momentum forms $T^i$: $F^i=8\pi GT^i$ (with $G=1$). The equations associated with $F^2$ and $F^3$ then yield two conditions. One is obtained from the real part 
\begin{equation} \label{v=-f}
   -\frac{e^{-2f}}{2r} \Bigg(\frac{\partial v}{\partial r}+ \frac{\partial f}{\partial r} \Bigg) =0 \quad  \Rightarrow \quad \frac{\partial v}{\partial r}=- \frac{\partial f}{\partial r},
\end{equation}
relating functions $v$ and $f$ by $v(r,t)=-f(r,t)$, and another condition from the imaginary part
\begin{equation} \label{fcnr}
     \frac{e^{-(f+v)}}{r}\frac{\partial f}{\partial t} =0 \quad  \Rightarrow \quad  \frac{\partial f}{\partial t}=0,
\end{equation}
saying that the function $f$ is independent of the variable $t$ and similarly $v(r,t) = v(r)$.
Finally, from $F^1$, we obtain the differential equation, which after using conditions (\ref{v=-f}) and (\ref{fcnr}) takes the form
\begin{equation} \label{RNode}
    e^{2v(r)}\bigg[r^2\frac{\partial^2 v}{\partial r^2}+2r^2\left(\frac{\partial v}{\partial r}\right)^2+e^{-2v(r)}-1\bigg]=\frac{2Q^2}{r^2}.
\end{equation}
The exact solution of the differential equation (\ref{RNode}) gives the Reissner-Nordstrom solution
\begin{equation} \label{soln}
  v(r) = \frac{1}{2}\ln{(1-\frac{2C_1}{r}+\frac{Q^2}{r^2}+2C_2r^2)},  
\end{equation}
where $C_1$ and $C_2$ are constants to be identified with the mass $M$ and cosmological constant $-\frac{\Lambda}{6}$, respectively.

The final trace equation can be found by taking the trace of the self-dual part of $F^i$. If we denote  the self-dual part of $F^i$ as $\Xi^{ij}$ which is the $3 \times 3$ self-dual/self-dual block of the Riemann tensor self-dual/anti-self-dual decomposition, then 
\begin{equation}
    \Xi^{ij} = \underbrace{(\Xi^{ij}-\frac{1}{3}\delta^{ij} \,Tr \Xi )}_{Trace-free \,Weyl \,Part} + \underbrace{\frac{1}{3}\delta^{ij} \,Tr  \Xi}_{Scalar \,Part}.
\end{equation}
So, taking the scalar part from $\Xi^{ij}$ and setting it equal to the scalar part of the Riemann decomposition ($\frac{R}{12}$) while considering the conditions (\ref{v=-f}) and (\ref{fcnr}) results in 
\begin{equation} \label{traceeq}
    \frac{1}{3}(-e^{2v}\left(\frac{\partial v}{\partial r}\right)^2-\frac{1}{2}e^{2v}\frac{\partial^2 v}{\partial r^2}-\frac{e^{2v}}{2r^2}+\frac{1}{2r^2}-\frac{2e^{2v}}{r}\frac{\partial v}{\partial r})= \frac{R}{12}
\end{equation}
Then simplifying (\ref{traceeq}) by substituting (\ref{soln}) gives in the known trace result
\begin{equation}
    R= 4 \Lambda.
\end{equation}

\section{Conclusion}
In this work we have revisited the introduction of energy-momentum in the Plebanski formulation of GR at the level of the field equations. Although the vacuum theory (namely the Einstein manifolds \cite{besse_einstein_1987}) have a particularly economical description in terms of the chiral soldering forms $\Sigma^i$ spanning $\Lambda^+$ and an $\text{SO}(3,\mathbb{C})$ connection $A^i$, the coupling of matter suffers from a loss of transparency because the space of symmetric 2-tensors are not immediately afforded a representation in terms of eigenspaces of the Hodge dual. The central aim of this paper was to make the transition from a given $T_{\mu \nu}$ to $T^i$ explicit through the representation theory of the algebraic space of curvature tensors. 

We have demonstrated that the anti-self-dual 2-form $T^i$ that encodes the trace-free energy-momentum tensor in the Plebanski equation (\ref{Pleb_Field_Eqn}) can be constructed through the use of the Kulkarni-Nomizu product,
\begin{equation*}
    T^i= 4i[P^+(\hat{T} \KN g)]^{0i}_{\;\; \mu \nu} dx^{\mu} \wedge dx^{\nu}.
\end{equation*}
The idea here is that one first lifts the trace-free irreducible component of $T_{\mu \nu}$ to $\mathcal{K}$ via $\hat{T} \KN g$. One then isolates the $(1,1)$ representation within this space (which corresponds to homomorphisms from self-dual to anti-self-dual 2-forms) via a projection into $\mathfrak{so}(3,\mathbb{C})$ indices, or adjoint components of the fundamental decomposition (\ref{Lie_Alg_Decomp}). Importantly, this expression also reproduces that introduced previously by Krasnov \cite{krasnov_formulations_2020, krasnov_plebanski_2011},
\begin{equation*}
    \big(4i\big[P^+ (\hat{T} \KN g)\big]^{0i}\big)_{\mu \nu} = \Sigma^{i \; \; \rho}_{[\mu} \hat{T}_{\nu]\rho},
\end{equation*}
verifying that this description is consistent.

In addition to the construction of $T^i$, we have confirmed that (\ref{TKNg}) is compatible with the standard formulation of energy-momentum conservation in the theory. Applying the chiral Bianchi identity $d^A F^i = 0$ to the Plebanski field equation with matter reduces to the standard conservation law 
\begin{equation*}
    \nabla_{\mu} T^{\mu \nu} = 0.
\end{equation*}
This is the central point of departure between GR and the unimodular theories: despite both admitting reformulations in terms of modified BF theories involving the Plebanski variables $\Sigma^i$ \cite{montesinos_trace-free_2025, gielen_unimodular_2024}, the latter has access to \textit{only} the $(1,1)$ sector of $\mathcal{K}$ through the field equations, while GR contains both trace and trace-free data. One has to instead assume the consequence of the chiral Bianchi identity in the unimodular set-up to recover the trace data necessary for equivalence to GR \cite{hughes_encoding_2026}.

Finally, we applied the construction to the electro-vacuum case. For the spherically symmetric electromagnetic stress-energy tensor, the KN expression gives a simple anti-self-dual source with only $T^1$ non-vanishing. Equating the anti-self-dual part of the curvature to this source recovers the expected restrictions on the metric functions and leads to the Reissner-Nordstrom-de Sitter solution. We have verified these results using the computational algebra package SageManifolds \cite{gourgoulhon_tensor_2015} and have provided an open access repository for working with this formalism which can be easily modified for various metric and energy-momentum structures \cite{joudyfjb_joudyfjbplebanski-formulation--energy-momentum_2026}. 

More broadly, the KN products serves as a representation theoretic bridge between the symmetric tensor language of the metric formalism and the chiral 2-form language of Plebanski gravity. It packages the energy-momentum tensor into the same algebraic curvature space in which the traditional orthogonal (Ricci) decomposition is normally understood, allowing for a specification of the components required by the self-dual/anti-self-dual splitting. This gives a concise and computable prescription for matter sources in the Plebanski formalism and may be useful for studying other physically important stress-energy tensors, as well as for clarifying the relation between chiral gravity, trace-free gravity and BF formulations.

\section{Acknowledgments}
Jack C. M. Hughes is grateful to Professor Glenn Muschert for his support.

\section{Funding}
This work received institutional funding from Khalifa University. Jack C. M. Hughes and Fedor V. Kusmartsev acknowledge support from the Khalifa University Research and Innovation Grant KU-INT-RIG-2024-046/8474000759. Joudy F. Jamal Beek and Fedor V. Kusmartsev acknowledge support from the Khalifa University Research and Innovation Grants KU-INT-RIG-2023-8474000546 and KU-INT-RIG-2024-8474000754. Fedor V. Kusmartsev also acknowledges support from the Thousand Talents Program and the President’s International Fellowship Initiative of the Chinese Academy of Sciences Awards.

\section{Conflict of Interest}
The authors of this work declare that they have no conflicts of interest.

\appendix

\section{Appendix A: Proof of Self-dual Projection Identity}
\label{prfP}
In this section, we prove the identity in equation (\ref{P+}). One can se also the proof from \cite{felice_relativity_1990} for a perspective that emphasizes spin variables. We first recall that 
\begin{equation} \label{P}
   P^{+ \; \rho\sigma}_{\mu \nu} = \frac{1}{2}\left( \delta^{\rho}_{[\mu} \delta^{\sigma}_{\nu]} - \frac{i}{2} \epsilon_{\mu \nu}^{\; \; \; \rho \sigma}\right)
\end{equation}
and define 
\begin{equation} \label{Sig+A}
    \Sigma^i_{\;\mu \nu}= i(e^0 \wedge e^i)_{\mu \nu}- \frac{1}{2}\epsilon^i_{\; jk}(e^j \wedge e^k)_{\mu \nu} = 2ie^0_{[\mu}e^i_{\nu]}-\epsilon^i_{\; jk}e^j_{[\mu}e^k_{\nu]},
\end{equation}
where $\epsilon^i_{\;jk}= -\epsilon^{0i}_{\;\;jk}$ is the structure constant for the Lie Algebra $\mathfrak{so}(3, \mathbb{C})$ which behaves like a Levi-Civita tensor in 3D.
We begin with 
\begin{equation}
\frac{1}{4}\Sigma^i_{\;\mu \nu} \Sigma_i^{\;\rho \sigma},
\end{equation}
and using the definition in (\ref{Sig+A}) while noting that $e^{0 \rho}=-e_0^{\rho}$ results in the four terms
\begin{equation}
   \underbrace{e^0_{[\mu}e^i_{\nu]}e_0^{[\rho}e_i^{\sigma]}}_A 
   \underbrace{- \frac{1}{2}i\epsilon_i^{\; mn}e^0_{[\mu}e^i_{\nu]}e_m^{[\rho}e_n^{\sigma]}}_B 
   \underbrace{+ \frac{1}{2}i\epsilon^i_{\; jk}e^j_{[\mu}e^k_{\nu]}e_0^{[\rho}e_i^{\sigma]}}_C
   \underbrace{+\frac{1}{4}\epsilon^i_{\; jk}\epsilon_i^{\; mn}e^j_{[\mu}e^k_{\nu]}e_m^{[\rho}e_n^{\sigma]}}_D
\end{equation}
which will be treated individually. Starting with term $A$, we first note that 
\begin{equation} \label{delts}
    \delta_{\mu}^{\rho} = t_{\mu}^{\rho} +s_{\mu}^{\rho},
\end{equation}
where $t_{\mu}^{\rho}= e_0^{\mu}e^0_{\rho}$ and $s_{\mu}^{\rho}= e_i^{\mu}e^i_{\rho}$.
Then expanding term A and using \ref{delts} gives
\begin{equation}
    e^0_{[\mu}e^i_{\nu]}e_0^{[\rho}e_i^{\sigma]} = \frac{1}{4}(t_{\mu}^{\rho}s_{\nu}^{\sigma} - t_{\mu}^{\sigma}s_{\nu}^{\rho} - t_{\nu}^{\rho}s_{\mu}^{\sigma} + t_{\nu}^{\sigma}s_{\mu}^{\rho}).
\end{equation}
Moving on to term $D$, we make use of the Levi-Civita tensor identity in 3D
\begin{equation}
    \epsilon^i_{\; jk}\epsilon_i^{\; mn} = 2\delta_j^{[m}\delta_k^{n]},
\end{equation}
and expand then simplify the term to get
\begin{equation}
    \frac{1}{4}\epsilon^i_{\; jk}\epsilon_i^{\; mn}e^j_{[\mu}e^k_{\nu]}e_m^{[\rho}e_n^{\sigma]} = \frac{1}{2}s_{[\mu}^{\rho}s_{\nu]}^{\sigma}.
\end{equation}
Adding terms $A$ and $D$ by substituting every $s_{\mu}^{\rho}= \delta_{\mu}^{\rho}-t_{\mu}^{\rho}$ results in 
\begin{equation}
    A +D = \frac{1}{2}\delta_{[\mu}^{\rho}\delta_{\nu]}^{\sigma}.
\end{equation}
We now use $\epsilon_i^{\;mn}= \epsilon_{0i}^{\;\;mn}$ to write term B as 
\begin{equation}
    - \frac{1}{2}i\epsilon_i^{\; mn}e^0_{[\mu}e^i_{\nu]}e_m^{[\rho}e_n^{\sigma]} = - \frac{1}{4}i\epsilon_{0i}^{\;\; mn}(e^0_{\mu}e^i_{\nu}-e^0_{\nu}e^i_{\mu})e_m^{\rho}e_n^{\sigma}.
\end{equation}
Similarly for term C, using $\epsilon^i_{\;jk}= -\epsilon^{0i}_{\;\;jk}$, it can be written as
\begin{equation}
   \frac{1}{2}i\epsilon^i_{\; jk}e^j_{[\mu}e^k_{\nu]}e_4^{[\rho}e_i^{\sigma]} =- \frac{1}{4}i\epsilon^{0i}_{\;\; jk}(e_0^{\rho}e_i^{\sigma}-e_0^{\sigma}e_i^{\rho})e^j_{\mu}e^k_{\nu}.
\end{equation}
Then using the following tetrad identity for the Levi-Civita tensor
\begin{equation}
    \epsilon_{\mu\nu}^{\;\;\; \rho\sigma}=\epsilon_{0i}^{\;\; jk}(e^0_{\mu}e^i_{\nu}-e^0_{\nu}e^i_{\mu})e_j^{\rho}e_k^{\sigma}+\epsilon^{\;\; 0i}_{jk}(e_0^{\rho}e_i^{\sigma}-e_0^{\sigma}e_i^{\rho})e^j_{\mu}e^k_{\nu}, 
\end{equation}
the terms $B$ and $C$ add up to give 
\begin{equation}
    B+C = - \frac{1}{4}i \epsilon_{\mu\nu}^{\;\;\; \rho\sigma}.
\end{equation}
Finally, putting all the terms together results in 
\begin{equation}
    A+B+C+D = \frac{1}{2}\delta_{[\mu}^{\rho}\delta_{\nu]}^{\sigma} - \frac{1}{4}i \epsilon_{\mu\nu}^{\;\;\; \rho\sigma},
\end{equation}
which is exactly the definition of the Self-dual Projection operator $P^{+ \; \rho\sigma}_{\mu \nu}$ in \ref{P}.
\section{Appendix B: Proof of Hodge dual Identity}
\label{prfH}
Here we prove the identity in equation (\ref{HD}). We begin with 
\begin{equation}
     \frac{1}{3!} \epsilon_{\gamma}^{\;\lambda \mu \nu} (3\nabla_{[\lambda}Y_{\mu \nu]}),
\end{equation}
and drop the anti-symmetrization brackets as the anti-symmetrization is captured by $\epsilon_{\gamma}^{\;\lambda \mu \nu}$, so it simplifies to
\begin{equation}
     \frac{1}{2} \epsilon_{\gamma}^{\;\lambda \mu \nu} (\nabla_{\lambda}Y_{\mu \nu}).
\end{equation}
 Then using $\nabla_{\lambda}\epsilon_{\gamma}^{\;\lambda \mu \nu} =0$ and relabeling $\lambda \rightarrow \sigma$, we get 
\begin{equation} \label{B1}
     \nabla^{\sigma}(\frac{1}{2} \epsilon_{\gamma \sigma}^{\;\; \; \mu \nu} Y_{\mu \nu}).
\end{equation} 
Finally, interchanging $\gamma$ with $\sigma$ and noticing that the action of the hodge dual on 2-forms is defined as $\star=\frac{1}{2} \epsilon_{\mu \nu}^{\;\; \; \rho \sigma}$ allows us to write (\ref{B1}) as
\begin{equation}
     -\nabla^{\sigma}(\star Y)_{\sigma \gamma}.
\end{equation}

\section*{References}
\bibliographystyle{iopart-num}
\bibliography{references}

\end{document}